# Crystal-field interaction and oxygen stoichiometry effects in strontium-doped rare-earth cobaltates

A. Furrer[1,*], A. Podlesnyak[2], M. Frontzek[1], I. Sashin[3], J.P. Embs[1], E. Mitberg[4], and E. Pomjakushina[5]

[1]Laboratory for Neutron Scattering, Paul Scherrer Institut, CH-5232 Villigen PSI, Switzerland
[2]Quantum Condensed Matter Division, Oak Ridge National Laboratory, Oak Ridge, TN 37831-6473, USA
[3]Joint Institute for Nuclear Research, 141 980 Dubna, Moscow Region, Russia
[4]Institute of Solid State Chemistry RAS, 620 219 Ekaterinburg GSP-145, Russia
[5]Laboratory for Developments and Methods, Paul Scherrer Institut, CH-5232 Villigen PSI, Switzerland

**Abstract**

Inelastic neutron scattering was employed to study the crystal-field interaction in the strontium-doped rare-earth compounds $R_xSr_{1-x}CoO_{3-z}$ (R=Pr, Nd, Ho, and Er). Particular emphasis is laid on the effect of oxygen deficiencies which naturally occur in the synthesis of these compounds. The observed energy spectra are found to be the result of a superposition of crystal fields with different nearest-neighbor oxygen coordination at the R sites. The experimental data are interpreted in terms of crystal-field parameters which behave in a consistent manner through the rare-earth series, thereby allowing a reliable extrapolation for rare-earth ions not considered in the present work.

*Corresponding author: albert.furrer@psi.ch

## I. INTRODUCTION

Rare-earth cobaltates of the perovskite type $RCoO_3$ (R=rare earth) have attracted much interest due to their rich phase diagrams involving structural, magnetic, spin-state and metal-insulator transitions [1-5] as well as due to their technological applications as oxidation catalysts, electrode materials and oxygen-permeable membranes [6-8]. More recent research focused on hole-doped cobaltates $R_xA_{1-x}CoO_3$ with A=Sr [9] and A=Ca [10] which, e.g., revealed the existence of spin-state polarons for R=La and A=Sr [11,12]. The magnetic properties of these compounds depend on the crystal-field and exchange interactions of the R and Co ions. While the interactions associated with the Co ions are well documented in the literature [13,14], not much is known about the interactions associated with the R ions. Often some numerical data are extrapolated from other perovskites such as $RNiO_3$ [15] and $RGaO_3$ [16,17], but this approach is highly questionable due to the variations of the charge distributions for different transition metal ions. In order to shed light on the crystal-field interaction at the R sites, we performed inelastic neutron scattering (INS) experiments for the compounds $R_xSr_{1-x}CoO_{3-z}$ (R=Pr, Nd, Ho, and Er; $0.10 \leq x \leq 0.15$; $0.23 \leq z \leq 0.30$).

The compound $SrCoO_3$ crystallizes in a simple cubic structure with space group *Pm-3m* [18]. For $R_xSr_{1-x}CoO_{3-z}$ the partial replacement of $Sr^{2+}$ ions by $R^{3+}$ ions induces spin-state transitions for part of the Co ions from high-spin $Co^{4+}$ to low-spin $Co^{3+}$ as well as ogygen deficiencies z. The latter lead to structural modifications which were examined in detail, *e.g.*, for R=Ho with x=0.1 and $0.15 \leq z \leq 0.49$ [19]. These details, however, are irrelevant for the present crystal-field study, since the local symmetry at the R sites remains essentially cubic, as verified in INS experiments performed for $Pr_{0.2}Sr_{0.8}CoO_{3-z}$ [20], $Ho_{0.1}Sr_{0.9}CoO_{3-z}$ [20], and $Pr_{0.5}Sr_{0.9}CoO_3$ [21]. The oxygen non-stoichiometry results in a superposition of crystal fields due to different oxygen coordinations at the $R^{3+}$ sites as most convincingly demonstrated e.g. for the perovskite cuprates $ErBa_2Cu_3O_{7-z}$ (0<z<1) [22]. The superposition of crystal fields produces both additional lines and line broadening in the observed energy spectra, thus a proper identification of lines is important for a reliable data analysis. In fact, the conclusions reached in INS experiments carried out so far for some $R_xSr_{1-x}CoO_{3-z}$ compounds [20,21] suffer from an incorrect line assignment and thereby from an incorrect parametrization of the crystal-field interaction. Here we show that these problems can be overcome by supporting model calculations of the oxygen stoichiometry effects as well as by a systematic investigation over the whole rare-earth series.

The present work is organized as follows. The experimental procedure is described in Sec. II, followed in Sec. III by a summary of the underlying theoretical background. The experimental results and their analyses are presented in Sec. IV. Finally, some conclusions are given in Sec. V.

## II. EXPERIMENTAL

### A. Sample synthesis

The $R_xSr_{1-x}CoO_{3-z}$ samples were prepared by a conventional solid state reaction using appropriate amounts of rare-earth oxides, $SrCO_3$, and $Co_3O_4$ with minimum purity of 99.99%. The oxygen deficiency z was determined by thermogravimetric hydrogen reduction with a precision of dz=±0.01 [23]. The single-phase character of the samples was confirmed by X-ray powder diffraction.

## B. INS experiments

The INS experiments were carried out with use of three spectrometers in order to cover different ranges of energy transfers ΔE. Well suited for large energy transfers is the inverted geometry time-of-flight (TOF) spectrometer KDSOG at the pulsed reactor IBR-2 of the Joint Institute for Nuclear Research, Dubna (Russia). The energy of the scattered neutrons was 4.8 meV as determined by Zn analyzers behind a Be filter cooled down to T=80 K, yielding an energy resolution of about 5% with respect to ΔE. The INS spectra were measured at seven scattering angles between 30º and 90º. The samples were placed into a closed-cycle dilution refrigerator yielding T>10 K. INS experiments at smaller energy transfers were carried out at the spallation neutron sources SINQ of the Paul Scherrer Institut, Villigen (Switzerland) and SNS of the Oak Ridge National Laboratory, Oak Ridge (USA) with use of the high-resolution TOF spectrometers FOCUS [24] and CNCS [25], respectively, yielding energy resolutions for neutron energy-loss processes of typically 1-2% with respect to ΔE. The samples were enclosed in Al cylinders (with 12 mm diameter and 45 mm height) and mounted in He cryostats to achieve temperatures T>1.5 K. Additional experiments were performed for the empty container as well as for vanadium to allow the correction of the raw data with respect to background, detector efficiency, absorption, and detailed balance according to standard procedures.

## III. THEORETICAL BACKGROUND

For the compounds $R_xSr_{1-x}CoO_{3-z}$ the dominant contribution to the crystal-field interaction at the $R^{3+}$ site (0,0,0) results from the nearest-neighboring oxygen ions located at positions (±a/2,±a/2,0), (±a/2,0,±a/2), and (0,±a/2,±a/2), where a is the lattice parameter. For z=0 all twelve positions are occupied by oxygen ions, but for z>0 some of these positions are vacant. There is a clear preference for the location of the oxygen vacancies as concluded from the observation of superstructures in the neutron diffraction patterns of $Ho_xSr_{1-x}CoO_{3-z}$ [19]. Oxygen is mainly lost from the basal planes of the octahedra around the cobalt ions, whereas the apical oxygen positions remain unaffected. Fig. 1 shows the fraction of different oxygen coordinations around the $R^{3+}$ ions as a function of the oxygen deficiency z. The results were obtained from a Monte Carlo study in a three-dimensional lattice of N×N×N unit cells with N=15. The Monte Carlo simulations were performed for 5≤N≤20, and convergence occurred for N≥10. Since all compounds investigated in the present work have oxygen deficiencies 0.2<z≤0.3, the deviations from the full twelve-fold oxygen coordination are substantial, so that the observed energy spectra have to be analyzed as a superposition of crystal fields with different oxygen coordination around the $R^{3+}$ sites.

For full twelve-fold oxygen coordination the crystal-field Hamiltonian has cubic point symmetry and is given by

$$H = B_4^0 O_4^0 + B_4^4 O_4^4 + B_6^0 O_6^0 + B_6^4 O_6^4 , \qquad (1)$$

where the four-fold symmetry axis is taken as polar axis. $B_n^m$ denote the crystal-field parameters and $O_n^m$ the Stevens operator equivalents [26]. The parameters $B_n^m$ can be described by the product of a charge term $A_n$ and a geometrical coordination factor $f_n^m$ defined by Hutchings [27]:

$$B_n^m = A_n f_n^m . \qquad (2)$$

For cubic symmetry we have the relations $B_4^4 = 5B_4^0$ and $B_6^4 = -21B_6^0$, thus we are left with only two independent fourth- and sixth-order crystal-field parameters in Eq. (1). Consequently the observation of two crystal-field transitions (i.e., their energies and intensities) is sufficient to unambiguously determine the crystal-field parameters, which can also be estimated on the basis of a simple point-charge model involving the nearest-neighboring oxygen coordination shell:

$$A_n = q \, |e| \, \chi_n \, \langle r^n \rangle . \tag{3}$$

Here, q is the charge of the oxygen ions, e the charge of the electron, $\chi_n$ a reduced matrix element [27], and $\langle r^n \rangle$ the nth moment of the radial distribution of the 4f electrons [28]. For (pseudo-)cubic perovskites, the parameters calculated from Eq. (3) usually predict the correct crystal-field ground state as well as the correct sequence of crystal-field states, apart from a scaling factor for the overall crystal-field splitting.

For $R^{3+}$ ions coordinated by less than twelve oxygen ions, the point symmetry is lowered and additional terms enter Eq. (1):

$$H = \sum_{m=0}^{2} B_2^m O_2^m + \sum_{m=0}^{4} B_4^m O_4^m + \sum_{m=0}^{6} B_6^m O_6^m . \tag{4}$$

The large number of independent crystal-field parameters makes the parametrization of crystal-field spectra somehow arbitrary. However, as repeatedly shown in the past for several perovskite compounds [15-17,29,30], the crystal-field parameters $B_n^m$ for a particular degree n can reasonably be determined by taking account of the nearest-neighboring coordination polyhedron:

$$B_n^m = B_n^0 \frac{f_n^m}{f_n^0}, \tag{5}$$

so that the number of independent crystal-field parameters in Eq. (4) is reduced to three. The geometrical coordination factors $f_n^m$ used in the present work are listed in Table I.

Eqs (3) and (5) are well suited to extrapolate the crystal-field parameters from a particular rare-earth compound with $R_1$ ions to another isostructural compound with $R_2$ ions through the relation [29]

$$A_n(R_2) = A_n(R_1) \frac{\chi_n(R_2) \cdot \langle r^n \rangle (R_2)}{\chi_n(R_1) \cdot \langle r^n \rangle (R_1)}. \tag{6}$$

As proven for the rare-earth series of copper-oxide perovskites [30], Eq. (6) predicts the fourth- and sixth-order crystal-field parameters with a precision of typically 10%.

Fig. 2 exemplifies the application of Eqs (1)-(5) for $Er_xSr_{1-x}CoO_{3-z}$ by comparing the crystal field at the $Er^{3+}$ site for twelve, eleven, and ten coordinating oxygen ions. The major effect of the latter two cases is the splitting of all the quartets $\Gamma_8^{(n)}$ into two doublets. Identical charge parameters $A_4$ and $A_6$ are used for all cases. The second-order charge parameter $A_2$, which vanishes for the case of full twelve-fold oxygen coordination, is seen to have little effect on the overall energy level sequence except for the splitting of the two lowest crystal-field states. In fact, this splitting turns out to be important for an unambiguous determination of $A_2$ from the experimental data, since Eq. (3) cannot be applied to second-order crystal-field parameters [29].

## IV. RESULTS AND DATA ANALYSIS

### A. $Er_xSr_{1-x}CoO_{3-z}$

The results of INS experiments performed for $Er_{0.15}Sr_{0.85}CoO_{2.71}$ with use of the TOF spectrometer KDSOG are shown in Fig. 3, together with data taken for the isostructural, nonmagnetic compound $La_{0.15}Sr_{0.85}CoO_{2.71}$, which demonstrates the absence of phonon scattering in the considered range of energy transfers. The $Er_{0.15}Sr_{0.85}CoO_{2.71}$ data are characterized by a strong asymmetric peak at $2<\Delta E<7$ meV and a double-peak structure at $10<\Delta E<20$ meV. Additional experiments were carried out for $Er_{0.10}Sr_{0.90}CoO_{2.70}$ at low energy transfers with use of the TOF spectrometer FOCUS as displayed in Fig. 4, which shows the existence of a partly resolved line peaking at $\Delta E \approx 0.3$ meV. All these features cannot be explained on the basis of the „ideal" crystal-field Hamiltonian (1), but the analysis of the observed energy spectra has to be based on a superposition of crystal fields with different oxygen coordination around the $Er^{3+}$ ions as outlined in Sec. III.

For an oxygen deficiency $z=0.29$ and a statistical distribution of the oxygen vacancies we expect the observed energy spectra to be a superposition of crystal fields resulting from 12-fold, 11-fold, and 10-fold oxygen coordination with approximate weights of 30%, 40%, and 20%, respectively (neglecting the small fraction of $Er^{3+}$ ions with 9-fold and 8-fold oxygen coordination, see Fig. 1). For 12-fold oxygen coordination we derive from Eqs (1) and (3) the quartet $\Gamma_8^{(3)}$ as the ground state, and the excited crystal-field states are − with increasing energy separation - the doublet $\Gamma_6$, the quartets $\Gamma_8^{(2)}$ and $\Gamma_8^{(1)}$, and the doublet $\Gamma_7$. The corresponding crystal-field level sequence, properly scaled to the observed energy spectra, is displayed in the middle of Fig. 2, which predicts the presence of two strong ground-state transitions $\Gamma_8^{(3)} \to \Gamma_6$ and $\Gamma_8^{(3)} \to \Gamma_8^{(2)}$, whereas the transitions to the two highest states $\Gamma_8^{(1)}$ and $\Gamma_7$ could not be resolved in the experiments because of their small transition probabilities. Statistically of equal importance are the crystal-field contributions for 11-fold and 10-fold oxygen coordination as displayed in Fig. 2. Obviously the observed low-energy line peaking at $\Delta E \approx 0.3$ meV (see Fig. 4) has to be associated with the ground-state doublet-doublet splittings a→b and a'→b', which fixes the second-order crystal-field parameter at $A_2=1.4(2)$ meV. This interpretation is supported by the temperature dependence of the intensities as shown in the insert of Fig. 4. The strong asymmetric peak at $2<\Delta E<7$ meV results then from the ground-state transitions $\Gamma_8^{(3)} \to \Gamma_6$, (a,b)→c, and (a',b')→c' with continuously decreasing energies, but a least-squares Gaussian fit allows a decomposition into only two lines (see Fig. 3). Moreover, for 10-fold coordination the model predicts a strong ground-state transition (a',b')→d' at $\Delta E \approx 7.5$ meV which, however, was not observed in the experiments (see Fig. 3). Obviously the oxygen vacancies are not statistically distributed, but there is a strong preference for 11-fold coordination. We therefore interpret the energy spectra as a superposition of crystal fields with 12-fold and 11-fold oxygen coordination. The double-peak structure at $10<\Delta E<20$ meV is then composed of three transitions which were least-squares fitted by Gaussian lines. The assignment of the transitions is indicated by arrows in Fig. 3. From the observed intensities and considering the transition probabilities shown in Fig. 2 we conclude that 25(5)% of the $Er^{3+}$ ions have an ideal 12-fold oxygen coordination, whereas 75(5)% of the $Er^{3+}$ ions are coordinated by eleven (and less) oxygen ions. The resulting crystal-field parameters are listed in Table II.

## B. $Ho_{0.10}Sr_{0.90}CoO_{2.73}$

Energy spectra observed for $Ho_{0.10}Sr_{0.90}CoO_{2.73}$ with use of the TOF spectrometers FOCUS and KDSOG are shown in Figs 5(a) and 5(b), respectively, which are complemented by high-resolution experiments performed on the TOF spectrometer CNCS displayed in Fig. 6. All these data are characterized by strong ground-state transitions centered at 1, 3, and 6 meV as well as by weak transitions around 15 and 20 meV, which were least-squares fitted by Gaussian lines as shown in Fig. 5.

For an oxygen deficiency z=0.27, the data analysis has to be based on a superposition of crystal fields with different oxygen coordination, similar to the case of $Er_xSr_{1-x}CoO_{2.71}$ discussed in Sec. IV.A. For 12-fold oxygen coordination the crystal-field states at the $Ho^{3+}$ site comprise four triplets $\Gamma_4^{(1)}$, $\Gamma_4^{(2)}$, $\Gamma_5^{(1)}$, and $\Gamma_5^{(2)}$, two doublets $\Gamma_3^{(1)}$ and $\Gamma_3^{(2)}$, and a singlet $\Gamma_1$, which are completely split into 17 singlet states for 11-fold oxygen coordination (and similarly for 10-fold oxygen coordination). However, these splittings could not by resolved in the experiments, but they give rise to a substantial broadening of the observed transitions as visualized in Figs 5 and 6. We therefore treat the observed energy spectra on the basis of Eq. (1) by taking account of the additional splittings as line broadening effects. The best agreement between the observed and calculated energy spectra is obtained by the crystal-field level sequence displayed on top of Fig. 5, and the corresponding crystal-field parameters are listed in Table II. The temperature dependence of the intensities of the lowest transition $\Gamma_5^{(1)} \rightarrow \Gamma_3^{(1)}$ supports this interpretation as shown by the insert in Fig. 6. It is interesting to note that the $\Gamma_5^{(1)} \rightarrow \Gamma_1$ transition, which is forbidden for cubic point symmetry, has gained intensity through the presence of $Ho^{3+}$ ions with 11-fold (and 10-fold) oxygen coordination.

## C. $Pr_{0.10}Sr_{0.90}CoO_{2.77}$

An energy spectrum observed for $Pr_{0.10}Sr_{0.90}CoO_{2.77}$ with use of the TOF spectrometer CNCS is shown in Fig. 7, which is characterized by partly resolved peaks at $\Delta E \approx 6$, 8, and 12 meV. The data were least-squares fitted by three Gaussian lines on top of a sloping background described by the power law $\alpha \cdot \exp(-\beta \cdot \Delta E)$. We extrapolated the crystal-field parameters from $Er_xSr_{1-x}CoO_{3-z}$ to $Pr_xSr_{1-x}CoO_{3-z}$ on the basis of Eq. (6) as listed in Table II. The resulting energy level scheme is displayed in Fig. 8(a), which predicts the lowest ground-state transition $\Gamma_5 \rightarrow \Gamma_3$ to be at $\Delta E = 11.3$ meV, i.e., exactly at the position of the highest-energy peak observed in Fig. 7. The double-peak structure at $\Delta E \approx 7$ meV can then be associated with crystal-field transitions for 11-fold oxygen coordination as shown in Fig. 8, where the second-order crystal-field parameter $A_2$ has to be adjusted to the experimental data. The best agreement is obtained for $A_2 = -20(4)$ meV as shown in Fig. 8(b). Excitations to crystal-field states above 20 meV could not be observed, which is most likely due to substantial line broadening resulting from the oxygen non-stoichiometry as well as to strong phonon scattering contributions at higher energy transfers [21].

## D. $Nd_{0.10}Sr_{0.90}CoO_{2.74}$

The crystal-field parameters extrapolated from $Er_xSr_{1-x}CoO_{3-z}$ to $Nd_xSr_{1-x}CoO_{3-z}$ on the basis of Eq. (6) are listed in Table II. They predict the quartet $\Gamma_8^{(1)}$ to be the ground state, and the excited states are the doublet $\Gamma_6$ at 17 meV and the quartet $\Gamma_8^{(2)}$ at 43 meV. However, INS experiments performed for energy transfers up to $\Delta E = 50$ meV with use of the TOF spectrometer CNCS failed to observe any intensity associated with crystal-field transitions for reasons given in Sec. IV.C; moreover, the first excited crystal-field transition $\Gamma_8^{(1)} \rightarrow \Gamma_6$ has a rather small transition probability. Nevertheless, high-resolution INS experiments carried out with use of the TOF spectrometers CNCS and FOCUS revealed the presence of a line at $\Delta E \approx 0.6$ meV as shown in Fig. 9. The magnetic origin of the line was confirmed by the Q-dependence of the intensity which was in agreement with the square of the magnetic form factor. Its origin is most likely due to the splitting of the ground-state quartet $\Gamma_8^{(1)}$ into two doublets as confirmed by the temperature dependence of the intensity shown in the insert of Fig. 9. The splitting probably results from a slight distortion of the cubic site symmetry around the $Nd^{3+}$ ions and can be described by the second-order crystal-field parameter $A_2$=-4.5(9) meV.

## V. DISCUSSION AND CONCLUSIONS

We presented a comprehensive INS analysis of the crystal-field interactions and associated oxygen stoichiometry effects in the $R_xSr_{1-x}CoO_{3-z}$ compounds. The systematic study including the heavy R ions Er and Ho as well as the light R ions Pr and Nd allowed a consistent parametrization in terms of phenomenological crystal-field parameters. The reliability of our analysis is supported by the rather good agreement of the effective charges of the nearest-neighbor coordinating oxygen ions for different R ions. Based on Eq. (3) we derive from the crystal-field parameters $A_4$ listed in Table II consistent oxygen charges q=-3.7(3)|e| and q=-3.4(3)|e| for the heavy R ions Er and Ho, respectively. The oxygen charges derived from the crystal-field parameter $A_6$ are equally consistent with q=-2.8(3)|e| and q=-2.4(1)|e| for Er and Ho, respectively. Deviations from the „ideal" oxygen charge q=-2|e| are expected due to the screening of the 4f electrons by the outer shells, which can be taken into account by adding in Eq. (3) shielding (and antishielding) factors defined in Refs [29,31]. For the light R ions Pr and Nd we failed to detect most of the excited crystal-field transitions due to the reasons explained in Sec. IV, but the observed low-energy features were found to be in good agreement with the crystal-field parameters extrapolated from R=Er.

The calculations describing the oxygen stoichiometry effects were based on Eqs (3) and (5) which correspond to a point-charge model involving only the nearest-neighbor oxygen coordination shell. This simple model turned out to work surprisingly well, as it was restricted to the short-range fourth- and sixth-order crystal-field parameters $A_4$ and $A_6$. On the other hand, the parametrization of the long-range second-order crystal-field term $A_2$ relied exclusively on particular features of the observed energy spectra.

Clementyev at al. [21] reported a significant change of the crystal-field potential upon variation of the Pr content x in the $Pr_xSr_{1-x}CoO_{3-z}$ compounds. The corresponding transitions observed in INS experiments are listed in Table III. By going from x=0.50 to x=0.10, the energy of the $\Gamma_5 \rightarrow \Gamma_3$ transition is drastically reduced from 15 meV to 11.3 meV, which cannot be understood by the small lattice expansion from a=3.83 Å to a=3.84 Å. However, as mentioned in Sec. I, the mutual replacement of $Pr^{3+}$ and $Sr^{2+}$ ions results in a change of the $Co^{4+}/Co^{3+}$ ratio, which obviously has substantial effects on the overall charge distribution and modifies the crystal-field potential accordingly.

Table III also supports our analysis of the observed energy spectra in terms of a superposition of crystal-fields with different oxygen coordination at the $R^{3+}$ sites. Practically all the $Pr^{3+}$ sites in $Pr_{0.50}Sr_{0.50}CoO_{2.98}$ are ideally coordinated by twelve oxygen ions, thus no lines associated with 11-fold (and 10-fold) oxygen coordination were observed below the $\Gamma_5 \rightarrow \Gamma_3$ transition [21]. On the other hand, oxygen non-stoichiometry effects dominate the crystal-field interactions in $Pr_{0.20}Sr_{0.80}CoO_{2.70}$. In fact, the line associated with the a→(b,d) transition turned out to be twice as strong as the $\Gamma_5 \rightarrow \Gamma_3$ transition [20], whereas for $Pr_{0.10}Sr_{0.90}CoO_{2.77}$ with an intermediate oxygen deficiency the intensities of the two transitions are roughly equal.

In conclusion, our work provides a consistent basis for the parametrization of the crystal-field potential in the $R_xSr_{1-x}CoO_{3-z}$ compounds, which was experimentally verified for R=Er and R=Ho, but it can reliably be extrapolated for the whole rare-earth series as proven for R=Pr and R=Nd. Moreover, particular emphasis was laid on oxygen non-stoichiometry effects which influence the crystal-field potential in a decisive manner, but they can conveniently be handled by the model calculations outlined in the present work.

## ACKNOWLEDGMENT

Part of this work was performed at the Swiss Spallation Neutron Source (SINQ), Paul Scherrer Institut (PSI), Villigen, Switzerland. Research at Oak Ridge National Laboratory's Spallation Neutron Source was supported by the Scientific User Facilities Division, Office of Basic Energy Sciences, US Department of Energy.

## References

[1] G. Zhang, E. Gorelov, E. Koch, and E. Pavarini, Phys. Rev. B **86**, 184413 (2012)

[2] Z. Jirák, J. Hejtmánek, K. Knížek, P. Novák, E. Šantavá, and H. Fujishiro, J. Appl. Phys. **115**, 17E118 (2014)

[3] B. Scherrer, A. S. Harvey, S. Tanasescu, F. Teodorescu, A. Botea, K. Conder, A. N. Grundy, J. Martynczuk, and L. J. Gauckler, Phys. Rev. B **84,** 085113 (2011)

[4] J. Yu, D. Phelan, and D. Louca, Phys. Rev. B **84**, 132410 (2011)

[5] J. A. Alonso, M. J. Martínez-Lope, C. de la Callea, and V. Pomjakushin, J. Mater. Chem. **16**, 1555 (2006).

[6] R. H. E. van Doorn and A. J. Burggraf, Solid State Ionics **128**, 65 (2000).

[7]  L. Malavasi, C. Tealdi, G. Flor, G. Chiodelli, V. Cervetto, A. Montenero, and M. Borell, Sens. Actuators B **105**, 407 (2005)

[8]  F. Capon, A. Boileau, C. Carteret, N. Martin, P. Boulet, and J. F. Pierson, J. Appl. Phys. **114**, 113510 (2013).

[9]  M. James, A. Tedesco, D. Cassidy, M. Colella, and P. J. Smythe, Journal of Alloys and Compounds **419**, 201 (2006).

[10] G. J. Thorogood, P.-Y. Orain, M. Ouvry, B. Piriou, T. Tedesco, K. S. Wallwork, J. Herrmann, and M. James, Solid State Sciences **13**, 2113 (2011).

[11] A. Podlesnyak, M. Russina, A. Furrer, A. Alfonsov, E. Vavilova, V. Kataev, B. Büchner, Th. Strässle, E. Pomjakushina, K. Conder, and D. I. Khomskii, Phys. Rev. Lett. **101**, 247603 (2008).

[12] A. Podlesnyak, G. Ehlers, M. Frontzek, A. Furrer, Th. Strässle, E. Pomjakushina, K. Conder, F. Demmel, and D. I. Khomskii, Phys. Rev. B **83**, 134430 (2011).

[13] D. Phelan, D. Louca, S. N. Ancona, S. Rosenkranz, H. Zheng, and J. F. Mitchell, Phys. Rev. B **79**, 094420 (2009)

[14] H. M. Aarbogh, J. Wu, L. Wang, H. Zheng, J. F. Mitchell, and C. Leighton, Phys. Rev. B **74**, 134408 (2006).

[15] S. Rosenkranz, M. Medarde, F. Fauth, J. Mesot, M. Zolliker, A. Furrer, U. Staub, P. Lacorre, R. Osborn, R. S. Eccleston, and V. Trounov, Phys. Rev. B **60**, 14857 (1999).

[16] A. Podlesnyak, S. Rosenkranz, F. Fauth, W. Marti, A. Furrer, A. Mirmelstein, and H. J. Scheel, J. Phys.: Condens. Matter **5**, 8973 (1993).

[17] A. Podlesnyak, S. Rosenkranz, F. Fauth, W. Marti, H. J. Scheel, and A. Furrer, J. Phys.: Condens. Matter **6**, 4099 (1994).

[18] H. L. Yakel, Jr., Acta Cryst. **8**, 394 (1955).

[19] S. Streule, M. Medarde, A. Podlesnyak, E. Pomjakushina, K. Conder, S. Kazakov, J. Karpinski, and J. Mesot, Phys, Rev. B **73**, 024423 (2006).

[20] A. Podlesnyak, A. Mirmelstein, N. Golosova, E. Mitberg, I. Leonidov, V. Kozhevnikov, I. Sashin, F. Altorfer, and A. Furrer, Appl. Phys. A **74**, S1746 (2002). The oxygen contents of the samples were indicated to be 3±0.005, but later re-examined to be 2.70(1) for Pr and 2.73(1) for Ho (private communication by A.P.).

[21] E. S. Clementyev, P. Alekseev, V. V. Efimov, I. O. Troyanchuk, A. S. Ivanov, V. N. Lazukov, and V. V. Sikolenko, Journal of Surface Investigation: X-ray, Synchrotron and Neutron Techniques **6**, 553 (2012). The exact oxygen content of the sample was 2.98 (private communication by E.S.C.).

[22] J. Mesot, P. Allenspach, U. Staub, A. Furrer, and H. Mutka, Phys. Rev. Lett. **70**, 865 (1993).

[23] K. Conder, E. Pomjakushina, A. Soldatov, and E. Mitberg, Mat. Res. Bull. **40**, 257 (2005).

[24] S. Janssen, J. Mesot, L. Holitzner, A. Furrer, and R. Hempelmann, Physica B **234-236**, 1174 (1997).

[25] G. Ehlers, A. A. Podlesnyak, J. L. Niedziela, E. B. Iverson, and P. E. Sokol, Rev. Sci. Instrum. 82, 085108 (2011).

[26] K. W. H. Stevens, Proc. Phys. Soc. A **65**, 209 (1952).

[27] M. T. Hutchings, in *Solid State Physics*, Vol. 16, edited by F. Seitz and D. Turnbull (Academic, New York, 1964), p. 227.

[28] W. B. Lewis, in *Magnetic Resonance and Related Phenomena*, edited by I. Ursu (Publishing House of the Academy of Romania, Bucharest, 1971), p. 717.

[29] A. Furrer and A. Podlesnyak, in *Handbook of Applied Solid State Spectroscopy*, edited by D.R. Vij (Springer, New York, 2006), p. 257.

[30] J. Mesot and A. Furrer, in *Neutron Scattering in Layered Copper-Oxide Superconductors*, edited by A. Furrer (Kluwer, Dordrecht, 1998), p. 335.

[31] R. M. Sternheimer, Phys. Rev. **146**, 140 (1966).

Table I. Geometrical coordination factors $f_n^m$ for twelve, eleven, and ten nearest-neighbor oxygen ions around the $R^{3+}$ site in the $R_xSr_{1-x}CoO_{3-z}$ compounds. The lattice parameter was set at a=3.84 Å. For the $f_n^m(11)$ values the particular choice of the vacant oxygen position is irrelevant. The $f_n^m(10)$ values were obtained from two vacant oxygen positions (±a/2,0,a/2). Other choices of vacant oxygen positions result in slight energy shifts of the order of 10%.

| n | m | $f_n^m(12)$ | $f_n^m(11)$ | $f_n^m(10)$ |
|---|---|---|---|---|
| 2 | 0 | 0 | -0.62440×10⁻² | -0.12488×10⁻¹ |
| 2 | 1 | 0 | 0.74928×10⁻¹ | 0 |
| 2 | 2 | 0 | -0.18732×10⁻¹ | -0.37464×10⁻¹ |
| 4 | 0 | -0.14821×10⁻² | -0.11380×10⁻² | -0.79396×10⁻³ |
| 4 | 1 | 0 | 0.10586×10⁻² | 0 |
| 4 | 2 | 0 | -0.26465×10⁻² | -0.52931×10⁻² |
| 4 | 3 | 0 | 0.74103×10⁻² | 0 |
| 4 | 4 | -0.74103×10⁻² | -0.83366×10⁻² | -0.92629×10⁻² |
| 6 | 0 | -0.13999×10⁻³ | -0.13147×10⁻³ | -0.12294×10⁻³ |
| 6 | 1 | 0 | -0.26384×10⁻³ | 0 |
| 6 | 2 | 0 | -0.23557×10⁻⁴ | -0.47114×10⁻⁴ |
| 6 | 3 | 0 | 0.47114×10⁻³ | 0 |
| 6 | 4 | 0.29399×10⁻² | 0.26855×10⁻² | 0.24311×10⁻² |
| 6 | 5 | 0 | 0.62190×10⁻³ | 0 |
| 6 | 6 | 0 | -0.51825×10⁻⁴ | -0.10365×10⁻³ |

Table II. Crystal-field parameters $A_n$ of the $R_xSr_{1-x}CoO_{3-z}$ compounds. The parameters $B_n^m = A_n f_n^m$ (Eq. 2) can be calculated with use of the geometrical coordination factors $f_n^m$ listed in Table I.

| Compound | $A_2$ [meV] | $A_4$ [meV] | $A_6$ [meV] | Remarks |
|---|---|---|---|---|
| $Er_{0.15}Sr_{0.85}CoO_{2.71}$ | 1.4(2) | 0.306(27) | $1.51(14) \times 10^{-2}$ | determined from Figs 3 and 4 |
| $Ho_{0.10}Sr_{0.90}CoO_{2.73}$ | | -0.233(19) | $-8.57(34) \times 10^{-3}$ | determined from Figs 5 and 6 |
| $Pr_{0.10}Sr_{0.90}CoO_{2.77}$ | | -14.2 | 1.79 | extrapolated from $Er_{0.15}Sr_{0.85}CoO_{2.71}$ |
| $Pr_{0.10}Sr_{0.90}CoO_{2.77}$ | -20(4) | | | determined from Fig. 7 |
| $Nd_{0.10}Sr_{0.90}CoO_{2.74}$ | | -4.7 | -0.89 | extrapolated from $Er_{0.15}Sr_{0.85}CoO_{2.71}$ |
| $Nd_{0.10}Sr_{0.90}CoO_{2.74}$ | -4.5(9) | | | determined from Fig. 9 |

Table III. Transitions observed in INS experiments performed for $Pr_xSr_{1-x}CoO_{3-z}$ as a function of the Pr content x. For the assignment of the transitions we refer to Fig. 8.

| Compound | $\Gamma_5 \to \Gamma_3$ | $a \to (b,d)$ | Reference |
|---|---|---|---|
| $Pr_{0.50}Sr_{0.50}CoO_{2.98}$ | 15 meV | - | 21 |
| $Pr_{0.20}Sr_{0.80}CoO_{2.70}$ | 13 meV | 8 meV | 20 |
| $Pr_{0.10}Sr_{0.90}CoO_{2.77}$ | 11.3 meV | 7 meV* | present work |

* corresponds to the average of the transitions $a \to b$ and $a \to d$ in Fig. 7

**Figure Captions**

FIG. 1. (Color online) Fraction of different oxygen coordinations around the $R^{3+}$ ions in $R_xSr_{1-x}CoO_{3-z}$ as a function of the oxygen deficiency z. The results were obtained from a Monte Carlo simulation in an N×N×N lattice of unit cells with N=15.

FIG. 2. (Color online) Crystal-field level sequence at the $Er^{3+}$ sites in $Er_xSr_{1-x}CoO_{3-z}$ for twelve, eleven, and ten coordinating $O^{2-}$ ions, calculated from Eqs (1)-(5) with $A_4$=0.306 meV and $A_6$=0.0151 meV. The arrows mark selected ground-state transitions, and the corresponding numbers denote the transition probability $\left|\left\langle \Gamma_n \left| J_\perp \right| \Gamma_m \right\rangle\right|^2$.

FIG. 3. (Color online) Energy spectra of neutrons scattered from $Er_{0.15}Sr_{0.85}CoO_{2.71}$ and $La_{0.15}Sr_{0.85}CoO_{2.71}$ at T=10 K. The data were taken with use of the TOF spectrometer KDSOG for moduli of the scattering vector 1.6<Q<3.3 Å$^{-1}$. The lines denote Gaussian fits as explained in the text. The arrows mark the crystal-field transitions according to Fig. 2.

FIG 4. (Color online) Energy spectra of neutrons scattered from $Er_{0.10}Sr_{0.90}CoO_{2.70}$. The data were taken with use of the TOF spectrometer FOCUS for moduli of the scattering vector 0.4<Q<2.2 Å$^{-1}$. The incoming neutron energy was 3.3 meV. The arrow marks the transition according to Fig. 2. The insert shows the temperature dependence of the intensities integrated in the range 0.2≤ΔE≤0.8 meV (spheres) and scaled to the calculated Boltzmann population factor for the ground-state doublet (line).

FIG. 5. Energy spectra of neutrons scattered from $Ho_{0.10}Sr_{0.90}CoO_{2.73}$. The lines denote Gaussian fits as explained in the text. The top displays the crystal-field level sequence derived from the data, and the arrows mark the observed transitions. (a) Data taken at T=1.5 K with use of the TOF spectrometer FOCUS for moduli of the scattering vector 0.5<Q<4.9 Å$^{-1}$. The incoming neutron energy was 15.5 meV. (b) Data taken at T=10 K with use of the TOF spectrometer KDSOG for moduli of the scattering vector 2.4<Q<3.7 Å$^{-1}$.

FIG 6. (Color online) Energy spectra of neutrons scattered from $Ho_{0.10}Sr_{0.90}CoO_{2.73}$. The data were taken with use of the TOF spectrometer CNCS for moduli of the scattering vector 0.8<Q<1.8 Å$^{-1}$. The incoming neutron energy was 3.31 meV. The arrows mark the transitions according to Fig. 5. The insert shows the temperature dependence of the intensity integrated in the range 0.8≤|ΔE|≤1.2 meV (spheres) and scaled to the calculated Boltzmann population factor for the ground-state triplet $\Gamma_5^{(1)}$ (full line) and the first excited doublet $\Gamma_3^{(1)}$ (broken line).

FIG. 7. (Color online) Energy spectrum of neutrons scattered from $Pr_{0.10}Sr_{0.90}CoO_{2.77}$ at T=3 K. The data were taken with use of the TOF spectrometer CNCS for moduli of the scattering vector 1.5<Q<2.5 Å$^{-1}$. The incoming neutron energy was 20 meV. The lines denote Gaussian fits on top of a power-law background as explained in the text. The arrows mark the crystal-field transitions according to Fig. 8.

FIG. 8. (Color online) Crystal-field level sequence at the $Pr^{3+}$ sites in $Pr_xSr_{1-x}CoO_{3-z}$ for twelve and eleven coordinating oxygen ions, calculated from Eqs (1)-(5) with $A_4$=-14.2 meV and $A_6$=1.79 meV. The rectangle in (a) is displayed as an enlarged figure in (b). The arrows mark selected ground-state transitions, and the corresponding numbers denote the transition probability $|\langle\Gamma_n|J_\perp|\Gamma_m\rangle|^2$.

FIG 9. (Color online) Energy spectrum of neutrons scattered from $Nd_{0.10}Sr_{0.90}CoO_{2.74}$ at T=1.5 K. The data were taken with use of the TOF spectrometer CNCS for moduli of the scattering vector 0.8<Q<1.8 Å$^{-1}$. The incoming neutron energy was 3 meV. The insert shows the temperature dependence of the intensity integrated in the range 0.35≤ΔE≤1.15 meV (spheres; data taken with use of the TOF spectrometer FOCUS) and scaled to the calculated Boltzmann population factor for the ground-state quartet $\Gamma_8^{(1)}$ with a doublet-doublet splitting of 0.6 meV (line).

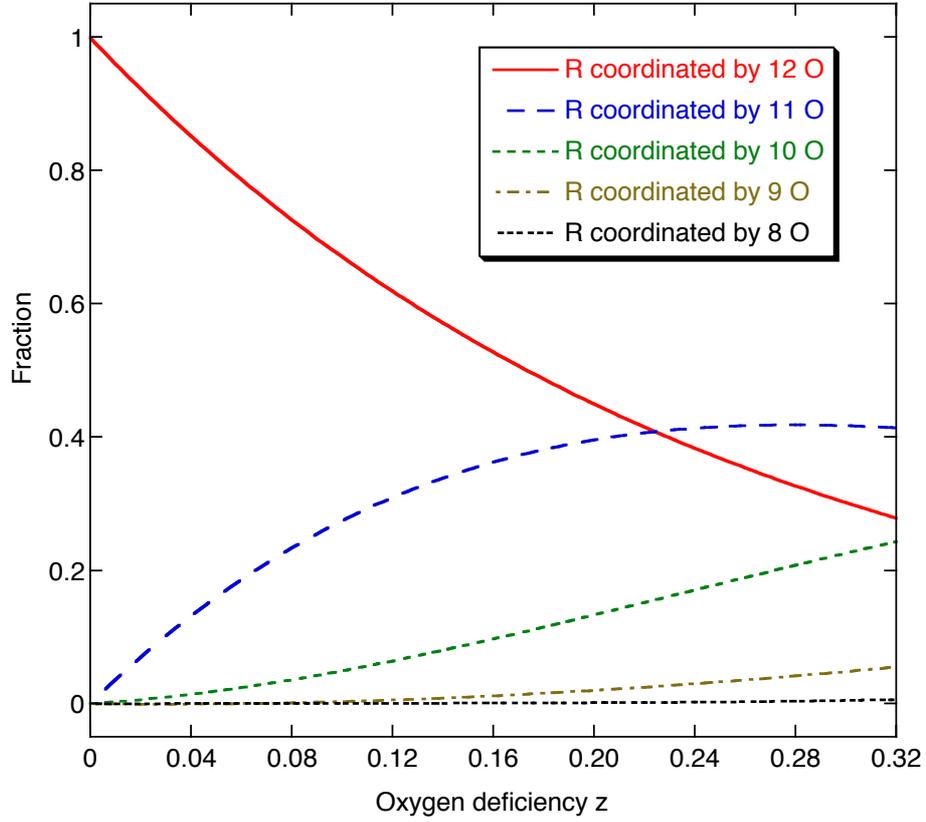

FIG. 1. (Color online) Fraction of different oxygen coordinations around the $R^{3+}$ ions in $R_xSr_{1-x}CoO_{3-z}$ as a function of the oxygen deficiency z. The results were obtained from a Monte Carlo simulation in an N×N×N lattice of unit cells with N=15.

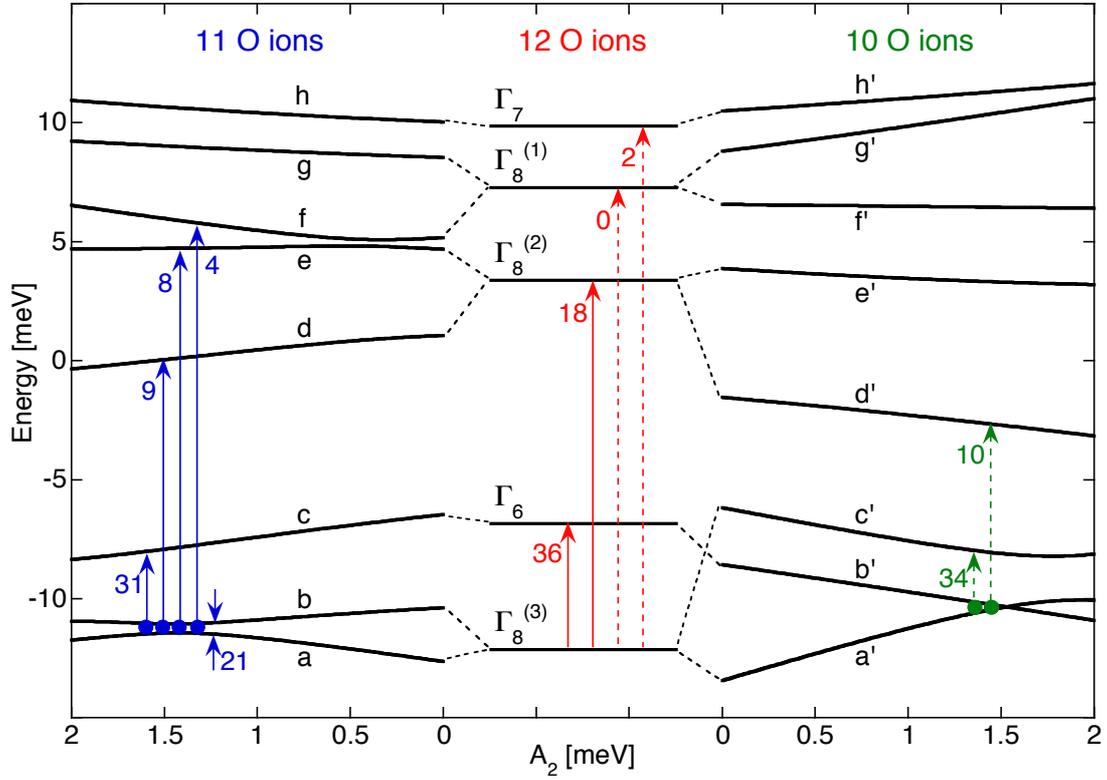

FIG. 2. (Color online) Crystal-field level sequence at the $Er^{3+}$ sites in $Er_xSr_{1-x}CoO_{3-z}$ for twelve, eleven, and ten coordinating $O^{2-}$ ions, calculated from Eqs (1)-(5) with $A_4$=0.306 meV and $A_6$=0.0151 meV. The arrows mark selected ground-state transitions, and the corresponding numbers denote the transition probability $|\langle \Gamma_n | J_\perp | \Gamma_m \rangle|^2$.

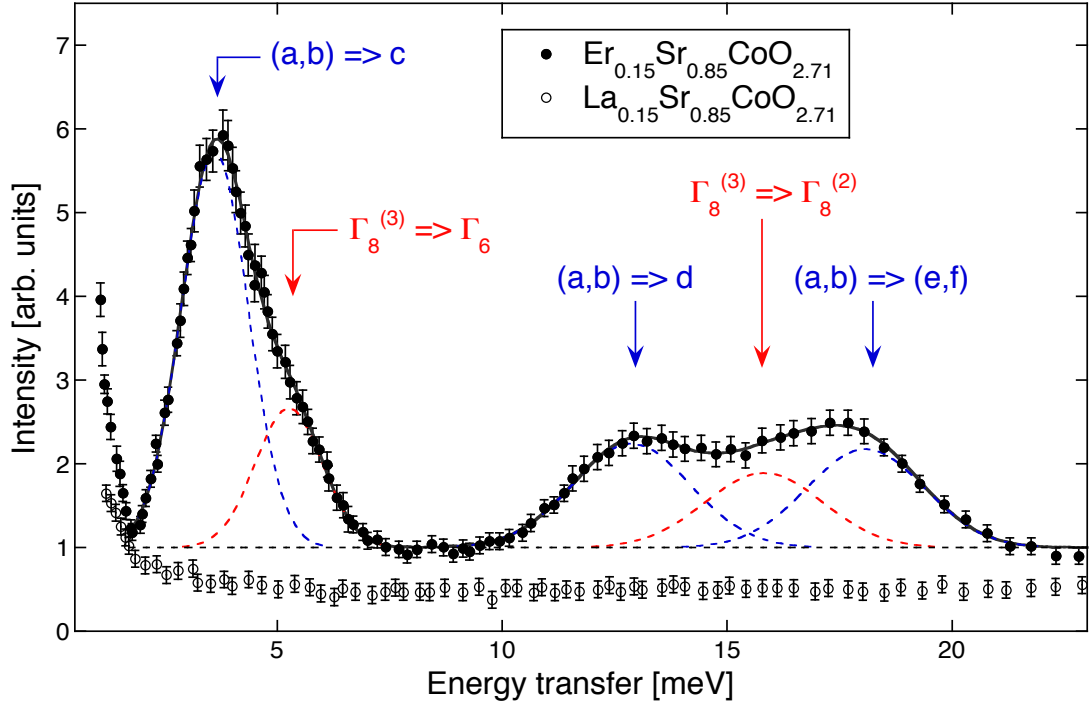

FIG. 3. (Color online) Energy spectra of neutrons scattered from $Er_{0.15}Sr_{0.85}CoO_{2.71}$ and $La_{0.15}Sr_{0.85}CoO_{2.71}$ at T=10 K. The data were taken with use of the TOF spectrometer KDSOG for moduli of the scattering vector 1.6<Q<3.3 Å$^{-1}$. The lines denote Gaussian fits as explained in the text. The arrows mark the crystal-field transitions according to Fig. 2.

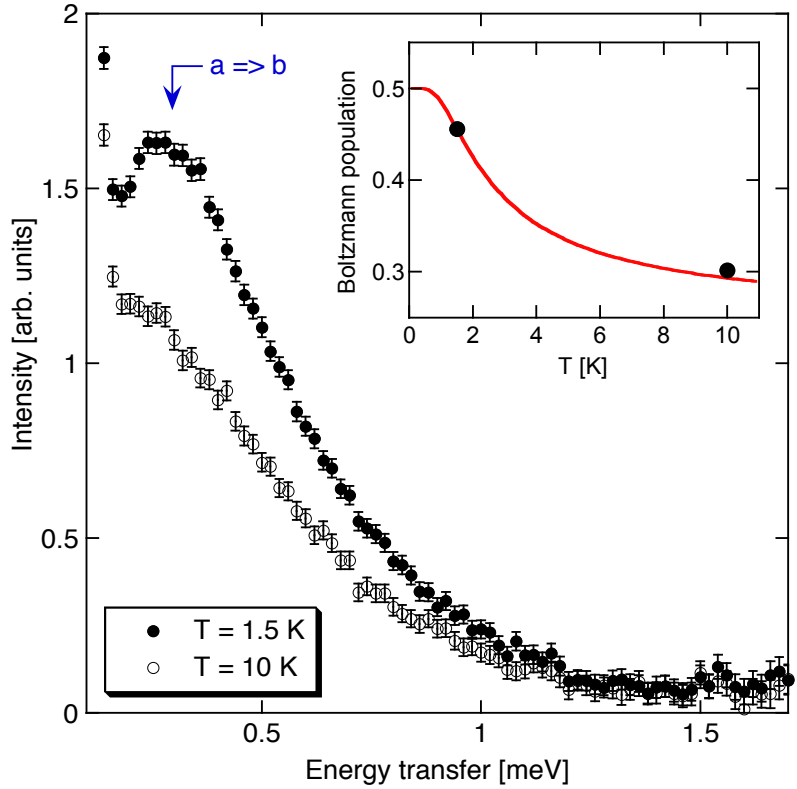

FIG 4. (Color online) Energy spectra of neutrons scattered from $Er_{0.10}Sr_{0.90}CoO_{2.70}$. The data were taken with use of the TOF spectrometer FOCUS for moduli of the scattering vector 0.4<Q<2.2 Å$^{-1}$. The incoming neutron energy was 3.3 meV. The arrow marks the transition according to Fig. 2. The insert shows the temperature dependence of the intensities integrated in the range 0.2≤ΔE≤0.8 meV (spheres) and scaled to the calculated Boltzmann population factor for the ground-state doublet (line).

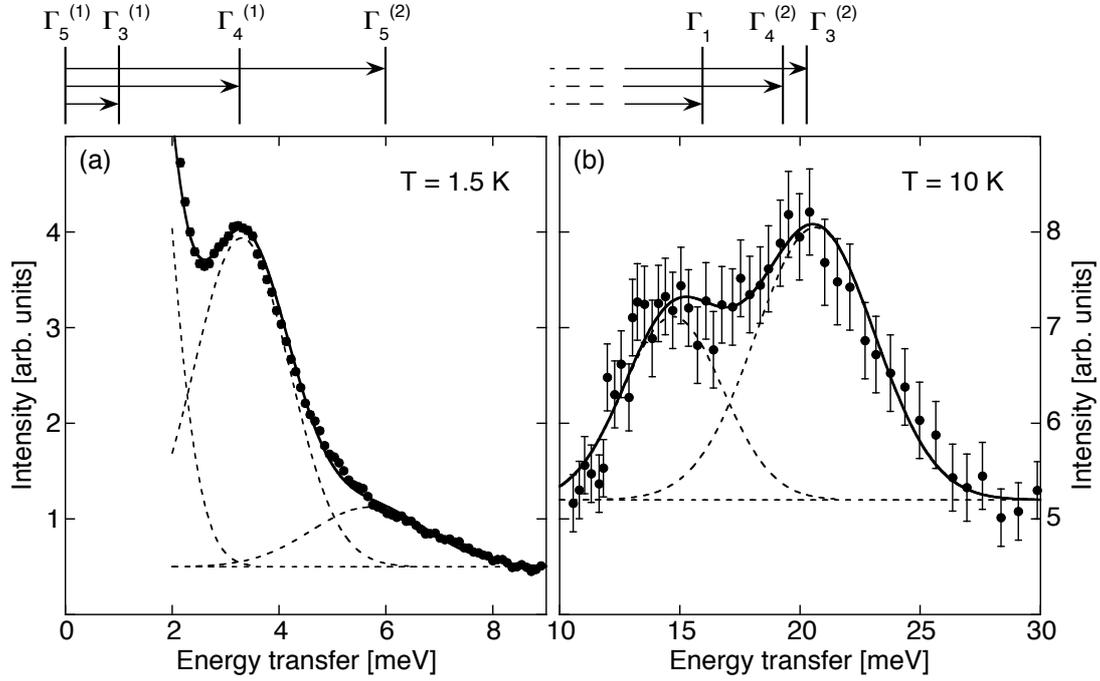

FIG. 5. Energy spectra of neutrons scattered from $Ho_{0.10}Sr_{0.90}CoO_{2.73}$. The lines denote Gaussian fits as explained in the text. The top displays the crystal-field level sequence derived from the data, and the arrows mark the observed transitions. (a) Data taken at T=1.5 K with use of the TOF spectrometer FOCUS for moduli of the scattering vector $0.5<Q<4.9$ Å$^{-1}$. The incoming neutron energy was 15.5 meV. (b) Data taken at T=10 K with use of the TOF spectrometer KDSOG for moduli of the scattering vector $2.4<Q<3.7$ Å$^{-1}$.

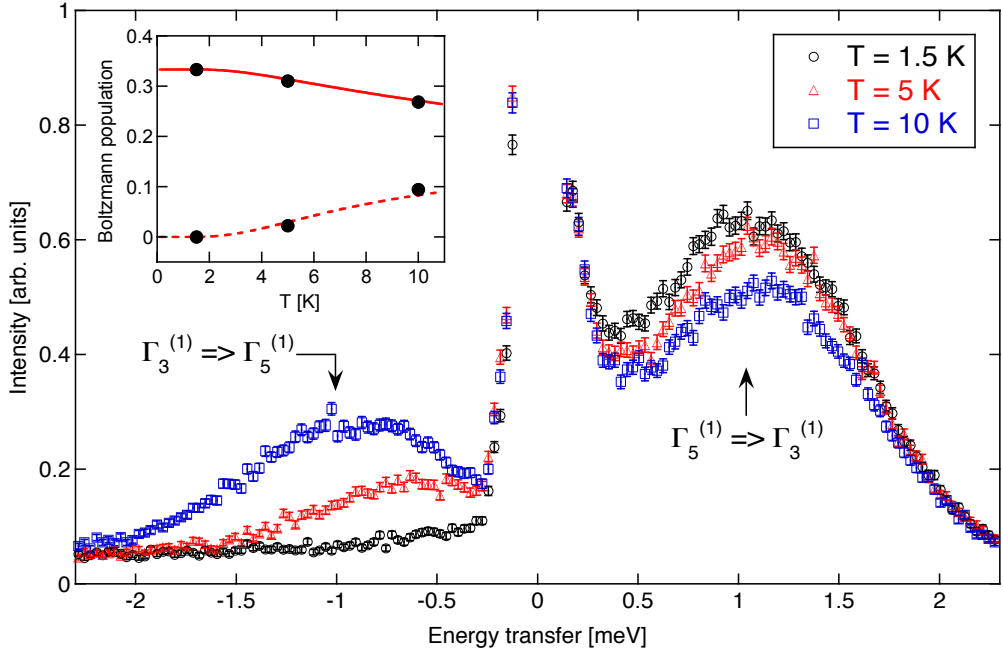

FIG 6. (Color online) Energy spectra of neutrons scattered from $Ho_{0.10}Sr_{0.90}CoO_{2.73}$. The data were taken with use of the TOF spectrometer CNCS for moduli of the scattering vector $0.8<Q<1.8$ Å$^{-1}$. The incoming neutron energy was 3.31 meV. The arrows mark the transitions according to Fig. 5. The insert shows the temperature dependence of the intensity integrated in the range $0.8\leq|\Delta E|\leq 1.2$ meV (spheres) and scaled to the calculated Boltzmann population factor for the ground-state triplet $\Gamma_5^{(1)}$ (full line) and the first excited doublet $\Gamma_3^{(1)}$ (broken line).

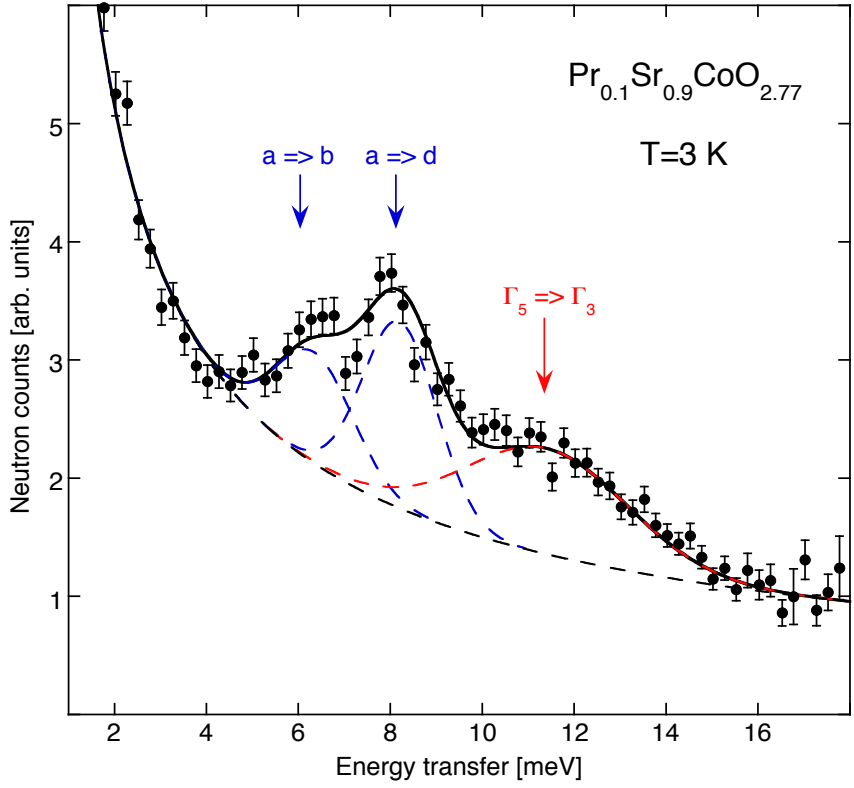

FIG. 7. (Color online) Energy spectrum of neutrons scattered from $Pr_{0.10}Sr_{0.90}CoO_{2.77}$ at T=3 K. The data were taken with use of the TOF spectrometer CNCS for moduli of the scattering vector $1.5<Q<2.5$ Å$^{-1}$. The incoming neutron energy was 20 meV. The lines denote Gaussian fits on top of a power-law background as explained in the text. The arrows mark the crystal-field transitions according to Fig. 8.

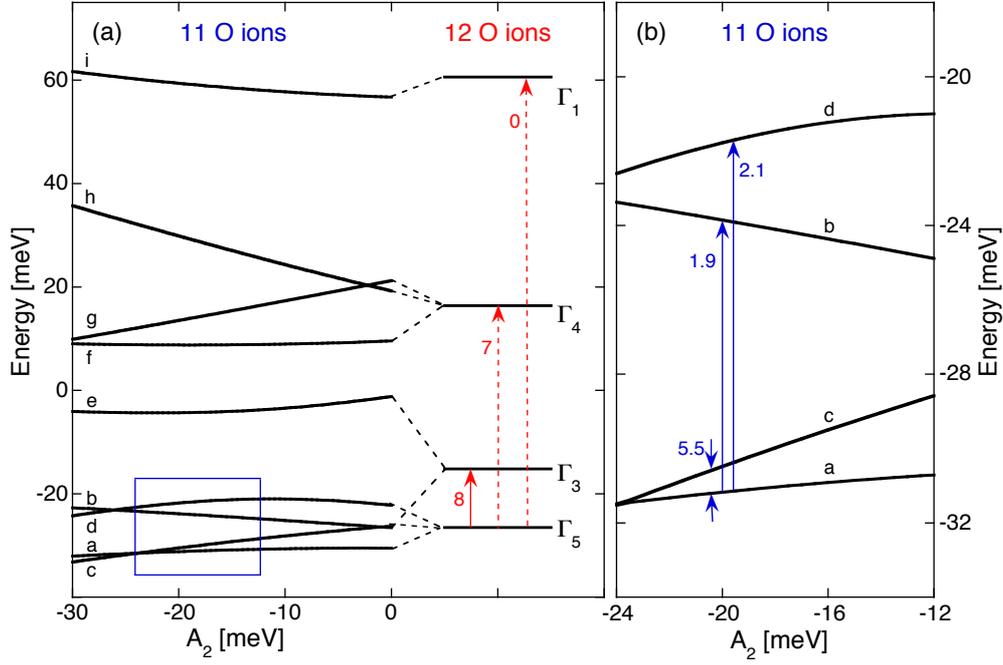

FIG. 8. (Color online) Crystal-field level sequence at the $Pr^{3+}$ sites in $Pr_xSr_{1-x}CoO_{3-z}$ for twelve and eleven coordinating oxygen ions, calculated from Eqs (1)-(5) with $A_4$=-14.2 meV and $A_6$=1.79 meV. The rectangle in (a) is displayed as an enlarged figure in (b). The arrows mark selected ground-state transitions, and the corresponding numbers denote the transition probability $|\langle \Gamma_n | J_\perp | \Gamma_m \rangle|^2$.

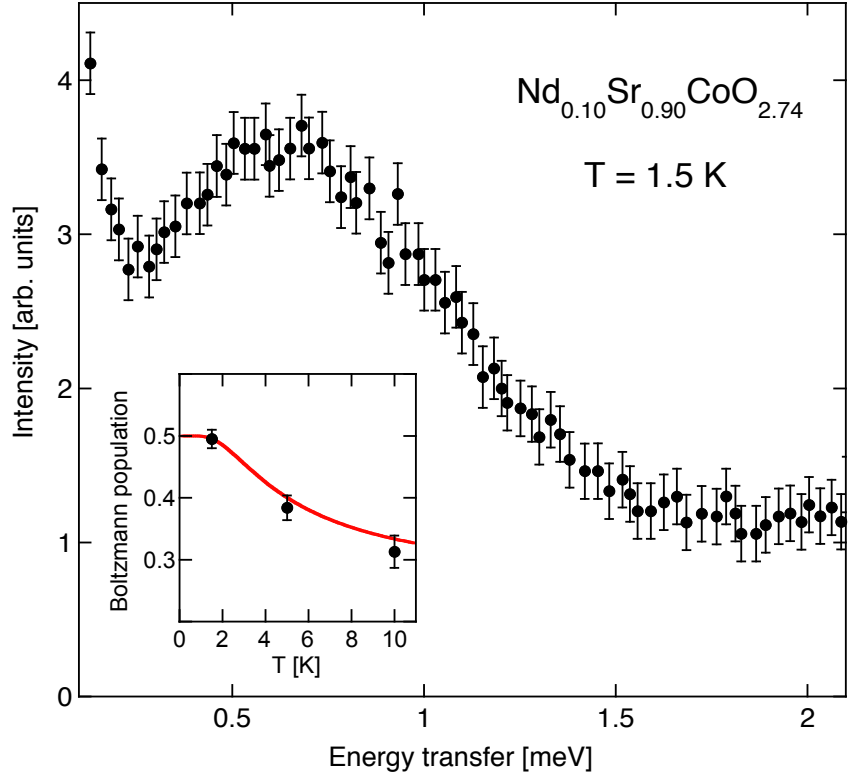

FIG 9. (Color online) Energy spectrum of neutrons scattered from $Nd_{0.10}Sr_{0.90}CoO_{2.74}$ at T=1.5 K. The data were taken with use of the TOF spectrometer CNCS for moduli of the scattering vector $0.8<Q<1.8$ Å$^{-1}$. The incoming neutron energy was 3 meV. The insert shows the temperature dependence of the intensity integrated in the range $0.35 \leq \Delta E \leq 1.15$ meV (spheres; data taken with use of the TOF spectrometer FOCUS) and scaled to the calculated Boltzmann population factor for the ground-state quartet $\Gamma_8^{(1)}$ with a doublet-doublet splitting of 0.6 meV (line).